\documentclass[english,showpacs,floatfix,prd]{revtex4}
\usepackage[utf8]{inputenc}
\usepackage[T1]{fontenc}
\usepackage{lmodern}
\usepackage{amsmath}
\usepackage{amssymb}
\usepackage{graphicx}
\usepackage{esint}
\usepackage{longtable}
\usepackage{babel} 

\begin{document}

\title{Deforming regular black holes}
\author{J. C. S. Neves}
\email{nevesjcs@ime.unicamp.br}
\affiliation{Instituto de Matemática, Estatística e Computação Científica, Universidade
Estadual de Campinas \\
 CEP. 13083-859, Campinas, SP, Brazil}

\begin{abstract}
In this work we have deformed regular black holes which possess a general mass term described by a function which generalizes the Bardeen and Hayward mass functions. By using linear constraints in the energy-momentum tensor to generate metrics, the solutions presented in this work are either regular or singular. That is, within this approach, it is possible to generate regular or singular black holes from regular or singular black holes. Moreover, contrary to the Bardeen and Hayward regular solutions, the deformed regular black holes may violate the weak energy condition despite the presence of the spherical symmetry. Some comments on accretion of deformed black holes in cosmological scenarios are made.
 \end{abstract}


\maketitle

\section{Introduction}
The existence of singularities is an open and interesting problem in both 
General Relativity (GR) and other theories of gravity. There is a common
belief that only a complete Quantum Theory of Gravity will solve
this problem. But without a fully developed candidate for a Quantum
Theory of Gravity, the singularities are avoided, for example, with
some violations in the energy conditions. And these violations are
more acceptable with the observation of cosmic acceleration \cite{Supernova,Supernova2}. By assuming these violations, the Hawking-Penrose theorems are not valid. In cosmology \cite{Novello} or black hole (BH) physics
\cite{Joshi}, we can create solutions of the gravitational field
equations without a singularity. Such solutions may violate the important
theorems of Hawking-Penrose and avoid the problem of singularities.

In Bardeen \cite{Bardeen2}, one has the first regular black hole
(RBH): a compact object with an event horizon and without a physical
singularity (see \cite{Ansoldi} for a review and \cite{Lemos_Zanchin} for a short introduction). This realization is a consequence of ideas from Sakharov and others \cite{Sakharov,Gliner} that the spacetime inside the horizon,
where the matter has a high pressure, is de Sitter-like. The Bardeen
solution is spherically symmetric and does not violate the weak energy
condition (WEC). According to \cite{Beato}, it is formed by a nonlinear
electrodynamic field. Decades later, Hayward \cite{Hayward} constructed
another regular solution with this symmetry. On the other hand, recently,
solutions with axial symmetry (BHs with rotation) have been developed
as well \cite{Various_axial,Various_axial2,Various_axial3,Various_axial4,Various_axial5,Neves_Saa}. In our work \cite{Neves_Saa}, we have
constructed a general class of regular solutions using the mass term
\begin{equation}
m(r)=\frac{M_{0}}{\left[1+\left(\frac{r_{0}}{r}\right)^{q}\right]^{p/q}},\label{Mass_term}
\end{equation}
where $M_{0}$ and $r_{0}$ are, respectively, mass and length parameters. To accommodate event horizon(s), one adopts $r_0 \ll M_0$ in (\ref{Mass_term}).
The Bardeen and Hayward BHs correspond to $p=3,q=2$ and $p=3,q=3,$
respectively. Our solution generalizes previous works because adopts
the cosmological constant, axial symmetry and the general mass term
(\ref{Mass_term}). Moreover, we show that the WEC is always violated
when the rotation is present. 

A general approach to build solutions in GR and brane
world contexts using linear constraints in the energy-momentum tensor is presented in \cite{Molina}. From known solutions (such as Schwarzschild, Reissner-Nordström etc.) it is possible to construct metrics deforming
the standard solutions. According to the constraint, it is possible
to obtain either regular or singular black holes, naked singularities
and wormholes. This approach generalizes results from refs. \cite{Molina2,Casadio,Bronnikov,Various,Various2,Various3} (such an approach has been useful to generate spherical RBHs in a brane world. But, in the axial case, RBHs with rotation need another strategy. For example, by using the Kerr-Schild ansatz \cite{Various4,Neves}). In this work, we shall adopt this procedure to construct solutions from the regular metric presented in \cite{Neves_Saa}. That is, our ansatz will have the
mass term given by eq. (\ref{Mass_term}). Then we shall show that from a regular BH is possible to obtain regular and singular BHs as well. 

The structure of this paper is as follows: in Section
2 we present the general features of RBHs with spherical symmetry;
in Section 3 we show the general approach to build solutions;
in Section 4 we apply this approach by using a general mass term;
in Section 5 we investigate the WEC violation; in Section 6
some comments on BHs in cosmological scenarios are made; in Section 7
the final remarks are presented.

In this work, we have used the metric signature $diag(-+++)$ and
the geometric units $G=c=1$, where $G$ is the gravitational constant,
and $c$ is the speed of light in vacuum.

\section{General features of spherical regular black holes}

According to \cite{Neves_Saa}, in eq. (\ref{Mass_term}), the only feasible value of $p$ is
3. That is, by adopting $p=3$, the metrics with the mass
function (\ref{Mass_term}), in the limit of small $r$, are de Sitter-like
(this result agrees with the comments made in Introduction on the interior region of RBHs, where there is a de Sitter core). The mass term function (\ref{Mass_term}) may be
written for small and large $r$, respectively, as
\begin{eqnarray}
m(r) & \approx & M_{0}\left(\frac{r}{r_{0}}\right)^{3},\label{Mass_approx_small}\\
m(r) & \approx & M_{0}\left(1-\frac{3}{q}\left(\frac{r_{0}}{r}\right)^{q}\right).\label{Mass_approx_large}
\end{eqnarray}
 It is worth noting that for small $r$, the mass function does not
depend on $q$. These approximations will be useful in the next sections. 

A family of RBHs with spherical symmetry,
\begin{equation}
ds^{2}=-A(r)dt^{2}+\frac{1}{B(r)}dr^{2}+r^{2}d\Omega^{2},\label{metric}
\end{equation}
where $d\Omega^{2}=d\theta^{2}+sin^{2}\theta d\phi^{2},$ may be written by using the mass function (\ref{Mass_term}). These metrics are not vacuum solutions of the gravitational field equations. The energy-momentum tensor, when $A(r)=B(r)=1-2m(r)/r$, is given by
\begin{equation}
T_{\nu}^{\mu}=\frac{1}{8\pi}\left(\begin{array}{cccc}
-\frac{2m'(r)}{r^{2}}\\
 & -\frac{2m'(r)}{r^{2}}\\
 &  & -\frac{m''(r)}{r}\\
 &  &  & -\frac{m''(r)}{r}
\end{array}\right),\label{Energy-momentum}
\end{equation}
(the symbol (') represents an ordinary derivative with respect to
$r$). For a general mass function $m(r)$\textemdash{}with the energy-momentum tensor in the form $T_{\nu}^{\mu}=diag\left(-\rho,p_{r},p_{t,}p_{t}\right)$, where $\rho,p_{r}$, and $p_{t}$ are the energy density, radial pressure
and tangential pressure, respectively\textemdash{}these metrics are
neither isotropic ($p_{r}=p_{t}$) nor have a traceless energy-momentum
tensor $(T_{\mu}^{\mu}=0)$. 

The standard RBHs (Bardeen and Hayward)
do not violate the WEC. That is, $\rho\geq0$ and $\rho+p_{i}\geq0$, for $i=r$ and
$i=t$, in these metrics. In these solutions, $A(r)=B(r)$ leads to $\rho+p_{r}=0$, $\rho=m'(r)/(4\pi r^2)$ and $\rho+p_{t}=(2m'(r)-rm''(r))/(8\pi r^{2})$, which are positive definite.
However, as we shall see, when $A(r)\neq B(r)$ this is not the case.
And WEC violations are possible even in the spherical case. On the other hand, as we have shown
in \cite{Neves_Saa}, the WEC violations are always present for RBHs in the axial case. 

The standard Bardeen and Hayward metrics are regular, i.e., the scalars are finite everywhere. For example, for $r\rightarrow 0$, the Bardeen and Hayward black holes have the Kretschmann scalar written as 
\begin{equation}
\lim_{r\rightarrow 0} K(r) = \lim_{r\rightarrow 0} R_{\mu\nu\alpha\beta}R^{\mu\nu\alpha\beta}=96\left(\frac{M_{0}}{r_{0}^3}\right)^2,\label{K_Bardeen}
\end{equation}
by using the notation of eq. (\ref{Mass_term}), where this limit is valid for any $q$ positive integer.    

In general, a metric in the form (\ref{metric}) has horizons given
by zeros of $A(r)$ and $B(r).$ Killing horizons are zeros of $A(r),$
and event horizons are roots of $B(r)$. With the mass term (\ref{Mass_term}),
RBHs present an outer horizon (the event horizon), $r_{+},$ and an
inner horizon, $r_{-}$, which are Killing horizons, $r_{k}$, only when
$A(r)=B(r)$. A static region is present outside of the event horizon
when $A(r)>0$ and $B(r)>0,$ where the Killing vector field $\xi_{t}=\frac{\partial}{\partial t}$ is timelike.

\section{The approach: deforming black holes}

In \cite{Molina} it is showed how to construct solutions in GR by means of deformations. By using a linear constraint
in the energy-momentum tensor, it is possible to obtain solutions which are deformations of the vacuum (or electrovacuum) standard solutions (Schwarzschild, Kerr, Reissner-Nordström etc) or standard RBHs. With the assumption that the field equations are
\begin{equation}
R_{\nu}^{\mu}-\frac{1}{2}R\delta_{\nu}^{\mu}=8\pi T_{\nu}^{\mu},\label{field_equations}
\end{equation}
the approach works in the brane world context as well \cite{Casadio,Bronnikov,Various,Various2,Various3,Lobo}.
In that context, by assuming vacuum on the brane, the "electrical" part of the five-dimensional Weyl tensor projected on the brane is interpreted as a traceless energy-momentum tensor: $E_{\mu\nu}=8\pi T_{\mu\nu}$,
with $E_{\mu}^{\mu}=0.$ Thus, in the brane world, the constraint,
from (\ref{field_equations}), is
\begin{equation}
R=0.\label{constraint_1}
\end{equation}
 Following the work \cite{Molina}, in GR context, we choose an isotropic
constraint as well:
\begin{equation}
p_{r}=p_{t},\label{constraint_2}
\end{equation}
which is important, for example, to build BH solutions in cosmological
scenarios (see \cite{Martin-Moruno,Neves2}). Another constraint in GR is given by
\begin{equation}
p_{r}=\omega\rho.\label{constraint_3-1}
\end{equation}
The constraint expressed in (\ref{constraint_3-1}) was used to build wormhole solutions in \cite{Lobo}. That is, the components of the energy-momentum tensor, $T^{\mu}_{\nu}=diag(-\rho,p_r,p_t,p_t)$, obey a linear constraint:
\begin{equation}
L(\rho,p_r,p_t)\equiv \alpha \rho +\beta p_r +2\gamma p_t =0. \label{constraint_0}
\end{equation} 
The constraint (\ref{constraint_1}) corresponds to $\alpha=-1,\beta=1$ and $\gamma=1.$ The isotropic
case\textemdash{}constraint (\ref{constraint_2})\textemdash{}is given
by $\alpha=0,\beta=1$ and $\gamma=-1/2,$ and the constraint (\ref{constraint_3-1})
is equivalent to $\alpha=\omega,\beta=-1$ and $\gamma=0$ in (\ref{constraint_0}). 

By using the linear constraint (\ref{constraint_0}) in the field equations (\ref{field_equations}), with aid of the metric (\ref{metric}), one has
\begin{equation}
\left(\frac{\alpha-\beta}{r^{2}}\right)\left[1-B(r)\right]+\frac{\gamma}{2}\left[\frac{A'(r)B'(r)}{A(r)}-\left(\frac{A'(r)}{A(r)}\right)^{2}\right]+\left(\frac{\beta+\gamma}{r}\right)\left[\frac{B(r)A'(r)}{A(r)}\right]-\left(\frac{\alpha-\gamma}{r}\right)B'(r)+\gamma\frac{A''(r)B(r)}{A(r)}=0.
\label{constraint_3}
\end{equation}
The general equation (\ref{constraint_3}) will be used to determine the elements of the metric given by eq. (\ref{metric}).

Deformation means: to obtain a set of solutions "close" to the known solutions. By choosing a function $A(r)=A_{0}(r)$, the approach yields to a function $B(r)$ derived from the standard case where $B(r)=A(r)$
in (\ref{metric}). The deformation is applied on the $B(r)$ component because its roots give, for example, the spacetime structure (event horizons when metric is written as eq. (\ref{metric}) and the physical singularity). Then, one seeks for new solutions, according to the spacetime structure, when one deforms the metric (\ref{metric}). In this work, we deal with deformations from the regular metric where
\begin{equation}
A(r)=A_{0}(r)=1-\frac{2m(r)}{r},\label{A_0}
\end{equation}
with $m(r)$ given by eq. (\ref{Mass_term}) (or (\ref{Mass_approx_small})
and (\ref{Mass_approx_large})). The ansatz is completed by
\begin{equation}
B(r)=A_{0}(r)+\left(C-1\right)B_{lin}(r),\label{B}
\end{equation}
where $B_{lin}(r)$ stands for the linear term of $B(r)$. The constant $C$ determines the type of solutions: for $C>0$
one has either regular or singular black holes, or naked singularities, $C<0$ (or $C<C_{0}$) may determine
wormholes, and for $C=C_{0}$ one has extremal black holes. For $C=1$
the standard cases are restored.

The function $B_{lin}(r)$ is determined from
\begin{equation}
B_{lin}(r)=A_{0}(r)\ exp\left[k\int\frac{dr}{rh(r)}\right],\label{Blin2}
\end{equation}
which is solution of eq. (\ref{constraint_3}) by using the ansatz (\ref{A_0})-(\ref{B}),
where 
\begin{equation}
h(r)=\left(\gamma-\alpha\right)A_{0}(r)+\frac{\gamma}{2}rA_{0}'(r).\label{h(r)}
\end{equation}
The constant $k$ is $-1$ when $p_{r}=p_{t}$, $k=-2$ when $R=0$, and $k=\omega+1$ when $p_{r}=\omega\rho$.

It is useful using the mass (\ref{Mass_approx_small}) and (\ref{Mass_approx_large})
to solve (\ref{Blin2}). Then, with the mass approximation for small $r$, eq. (\ref{Blin2}) reads
\begin{equation}
B_{lin}^{small}(r)=A_{0}(r)\left[\frac{r}{\left(2M_{0}(\alpha-2\gamma)r^{2}-(\alpha-\gamma)r_{0}^{3}\right)^{\frac{1}{2}}}\right]^{\left(\frac{k}{\gamma-\alpha}\right)}.\label{B_small}
\end{equation}
The denominator of (\ref{B_small}) is $h(r)r_{0}^{3}$. On the other
hand, for large $r$, we use rational functions for any $q>0$ to solve $B_{lin}(r)$, i.e., 
\begin{equation}
B_{lin}^{large}(r)=A_{0}(r)\prod_{n=1}^{q+1}\frac{1}{\vert r-r_{n}\vert^{\left(\frac{kc_{n}}{\alpha-\gamma}\right)}},\label{B_large}
\end{equation}
with
\begin{equation}
c_{n}=r_{n}^{q}\ \prod_{i=n+1}^{q+1}\prod_{\stackrel{j=1}{\stackrel{0<j<n}{}}}^{n-1}\frac{1}{\left(r_{n}-r_{i}\right)\left(r_{n}-r_{j}\right)}.\label{cn}
\end{equation}
The $r_{n}$ are the roots of $h(r)$. In general, $h(r)$ has $q+1$
roots, using the mass function (\ref{Mass_approx_large}) and since
$(q-1)\gamma+2\alpha\neq0$. In the regime that we shall adopt henceforth, $q=2$
and $q=3$ (Bardeen and Hayward solutions, respectively), this function has two positive real roots, $r_{0-}<r_{0+}$. 

In the next section, we shall use a regular ansatz to construct/deform solutions.

\section{Deforming a regular black hole}

Because of the regularity of $A_{0}(r)$, we shall focus on $B_{lin}(r)$. The ansatz in this paper has the form (\ref{A_0}) with $p=3$ in the mass function. Due to the length or to the cumbersome expressions,
the functions $B_{lin}(r)$ (for either small or large $r's$) shown
below were obtained by using eqs. (\ref{Mass_approx_small}) and (\ref{Mass_approx_large}).
However, in the Figures, we have used $m(r)$ given by (\ref{Mass_term})
to generate $B(r)$.

\subsection{Isotropic case, $p_{r}=p_{t}$}

When $\alpha=0,\beta=1,$ $\gamma=-1/2$, and $k=-1$, the eq. (\ref{B_small})
admits the following solution for small $r$:
\begin{equation}
B_{lin}^{small}(r)=\frac{A_{0}(r)\ r^{2}}{\left(2M_{0}r^{2}-\frac{r_{0}^{3}}{2}\right)}.\label{Blin_isotropic_small}
\end{equation}
This function diverges at $r=r_{0-}=\frac{1}{2}\sqrt{r_{0}^{3}/M_{0}}$,
which corresponds to the smaller positive real root of $h(r)$. Thus,
for $C\neq1,$ all metrics diverge, and the Kretschmann scalar simply reads
\begin{equation}
K\sim\frac{1}{\left(4M_{0}r^2-r_{0}^3\right)^{4}}.\label{Kretschmann_iso}
\end{equation}
This result is valid for any value of $C\neq1$ in (\ref{B})
for the isotropic case. On the other hand, $B(r)$ and $K$ do not diverge at $r=r_{0+}$ (the largest positive real root of $h(r)$) because the value of $c_{0+}$ is positive in (\ref{B_large}).

To study the spacetime  structure, we need to know the localization
of the event and Killing horizons. The latter correspond to solutions
of $A_{0}(r_{k_{i}})=0$, and the inner and the outer horizon (an event horizon) are
located at $r_{-}$ and $r_{+}$, respectively, which are roots of
$B(r_{\pm})=0$. According to the value of $C$, the number of zeros of
$B(r)$ is different (see Fig. 1). In general, assuming the existence
of a static region, $A(r)$ and $B(r)$ must be positive definite
for $r>r_{+}.$ Therefore, this assumption eliminates values of $C\leq0.$ 

For $C>1$ the function $B(r)$ has two zeros, $r_{-}<r_{+}$. The
largest zero corresponds to the event horizon $r_{+}$. The
metric is singular at $r=r_{0-}<r_{-}$. Then, the spacetime  structure
reads
\begin{equation}
r_{0-}<r_{-}=r_{k_{1}}<r_{+}=r_{k_{2}}<\infty.\label{structure_1}
\end{equation}
This case corresponds to the singular BH in a noncompact universe,
such as presented in \cite{Molina2}, which uses the same ansatz but with $m$
constant (our value of $C$ corresponds to the $C+1$ in the cited
article). 

When $0<C<1$ the situation is quite different: there are three zeros
in $B(r)$ (the third zero is not zero of $A(r)$). There is a maximum
value of $r$ $(r_{max}>r_{+})$. Such as \cite{Molina2}, these values
of $C$ lead to a singular BH in a compact universe. The spacetime
 structure is given by
\begin{equation}
r_{0-}<r_{-}=r_{k_{1}}<r_{+}=r_{k_{2}}<r_{max},\label{structure_2}
\end{equation}
 with
\begin{equation}
\lim_{C\rightarrow1}r_{max}=\infty.\label{rmax}
\end{equation}

These results are the same of ref. \cite{Molina2}. The difference is the
localization of the physical singularity: with the mass function (\ref{Mass_term})
or (\ref{Mass_approx_small}), the singularity is translated to $r=r_{0-}$.
Unfortunately the ansatz (\ref{A_0}) does not prevent the occurrence of the physical
singularity for the isotropic deformed geometries $(C\neq1)$.

\begin{figure}
\begin{centering}
\includegraphics[scale=0.47]{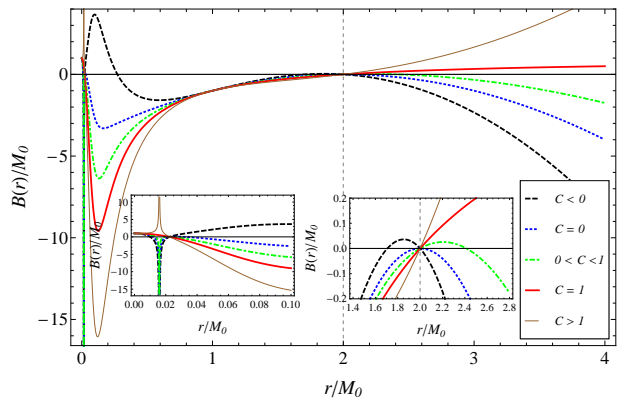}
\par\end{centering}

\caption{$B(r)$ obtained with the mass function (\ref{Mass_term}) and the
constraint $p_{r}=p_{t}$. The vertical dashed line corresponds to the event
horizon $r_{+}$. In this graphic we have used $r_{0}/M_{0}=0.1$, $q=3$, $C=-1$, $C=0.5$, and $C=2$. }

\end{figure}

\subsection{Brane world or traceless energy-momentum tensor, $R=0$ }

When $\alpha=-1,\beta=1,\gamma=1,$ and $k=-2$, the constraint $R=0$,
for small $r$, admits the following solution for (\ref{B_small}):
\begin{equation}
B_{lin}^{small}(r)=\frac{A_{0}(r)}{r}\sqrt{-6M_{0}r^{2}+2r_{0}^{3}}.\label{Blin_brane_small}
\end{equation}
$B_{lin}^{small}(r)$ diverges only at $r=0$. However, there is a
singular point for large $r$. That is, the function $h(r)$ has a
root $r_{0+}$ and the exponent $kc_{0+}/(\alpha-\gamma)$, in eq.
(\ref{B_large}), is positive. Therefore, $B(r)$ diverges at this
point (see Fig. 2). Typically, $r_{0+}\approx3M_{0}/2.$ This point may
indicate the physical singularity depending on the value of $C$.
Then, in the case $C\neq1,$ the Kretschmann scalar will read 
\begin{equation}
K(r)\sim\frac{1}{\left(r-r_{0+}\right)^{4}}.\label{Kretschmann_brane}
\end{equation}

The family of solutions generated from the regular ansatz (\ref{A_0}) in the case where $R=0$ has the same behaviour illustrated in \cite{Casadio,Bronnikov}. Again, when $C>1$ we have singular
BHs, where the physical singularity is located at $r_{0+}$; $C=C_{0}$
fixes the extremal case, the root $r=r_{+}$ is double; and $C<C_{0}$
generates a wormhole, with $r_{thr}$ indicating the throat of the
hole. Lastly, the case $C_{0}<C<1$ presents a regular black hole, where there
is a minimum value of $r=r_{min}$: this point is the end of the spacetime,
and the Kretschmann scalar is always finite because $r_{min}>r_{0+}$.
The latter two cases ($C<C_{0}$ and $C_{0}<C<1$) are possible because $B(r)$ has a second root.

For $C>1$ and $C=C_{0}$, the spacetime  structure is given by
\begin{equation}
r_{0+}<r_{k}=r_{+}<\infty.\label{structure_3}
\end{equation}
 When $C<C_{0},$ the radial coordinate assumes $[r_{thr},\infty[$, i.e., the second root of $B(r)$ in this case is $r_{thr}>r_k$, and the spacetime is static only when $r\geq r_{thr}$. The spacetime  structure in the single regular case $C_{0}<C<1$ reads
\begin{equation}
r_{min}<r_{k}=r_{+}<\infty.\label{structure_4}
\end{equation}

\begin{figure}
\begin{centering}
\includegraphics[scale=0.475]{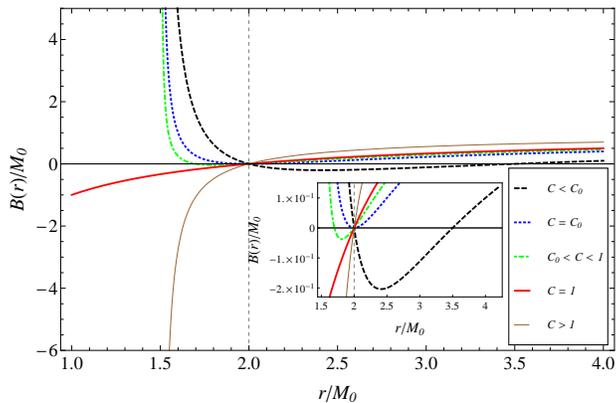}
\par\end{centering}

\caption{$B(r)$ obtained with the mass function (\ref{Mass_term}) and the
constraint $R=0$. The vertical dashed line corresponds to the event horizon
$r_{+}$. In this graphic we have used $r_{0}/M_{0}=0.1$, $q=3$, $C=-1$, $C_{0}=0.5$, $C=0.8$, and $C=2$.}

\end{figure}

\subsection{Constraint $p_{r}=\omega\rho$}

When $\alpha=\omega,\beta=-1,\gamma=0$ and $k=\omega+1$, the function
$h(r)$ is quite simple, by using the mass function (\ref{Mass_term})
or (\ref{Mass_approx_small})-(\ref{Mass_approx_large}). In this
case, $h(r)=-\omega A(r)$, and the divergence of $B(r),$ for the
deformed metrics, occurs in either $r=r_{0-}=r_{k_{1}}$ or $r=r_{0+}=r_{k_{2}}$.
For large $r$, one has 
\begin{equation}
B_{lin}^{large}(r)\sim\frac{1}{\vert r-r_{k_{2}}\vert^{\left(c_{0+}\left(\frac{1+\omega}{\omega}\right)-1\right)}},\label{Blin_worm_large}
\end{equation}
with $c_{0+}\approx1$ (the minus one in the exponent comes from $A_{0}(r)\sim(r-r_{k_{2}})$).
Then, the divergence of $B(r)$ in $r=r_{k_{2}}$ depends on the sign
of $\omega.$ But the divergence of the Kretschmann scalar in this
point, due to the existence of this root for $h(r),$ is not given
by the divergence of \textbf{$B(r).$} The Kretschmann scalar reads
\begin{equation}
K(r)\sim\frac{\left(\left(1+\omega\right)c_{0+}\right)^{2}}{\omega^{2}(r-r_{k_{2}})^{2\left(1+\left(\frac{1+\omega}{\omega}\right)c_{0+}\right)}}.\label{Kretschmann_case3}
\end{equation}
 Therefore, this metric has a physical singularity at $r_{k_{2}}$
when $(1+\left(\frac{1+\omega}{\omega}\right)c_{0+})>0$ $(\omega>0$
or $\omega<-c_{0+}/(1+c_{0+}))$. The behaviour of $B(r)$ in this
case is shown in Fig. 3 (left). For $C>1$, we have a naked
singularity. There is no event horizon or, equivalently,
a root of $B(r)$ in such a spacetime. With $C<1,$ the geometries are described by wormholes.
There is a large root for $B(r_{thr})$ such that $r_{thr}>r_{0+}=r_{k_{2}}$,
and the metric signature is incorrect for $r<r_{thr}$. Thus, for
$(1+\left(\frac{1+\omega}{\omega}\right)c_{0+})>0,$ the physical
solutions considered here are the standard solution $(C=1)$, naked
singularities $(C>1)$ and wormholes $(C<1)$. 

\begin{figure}
\begin{centering}
\includegraphics[scale=0.43]{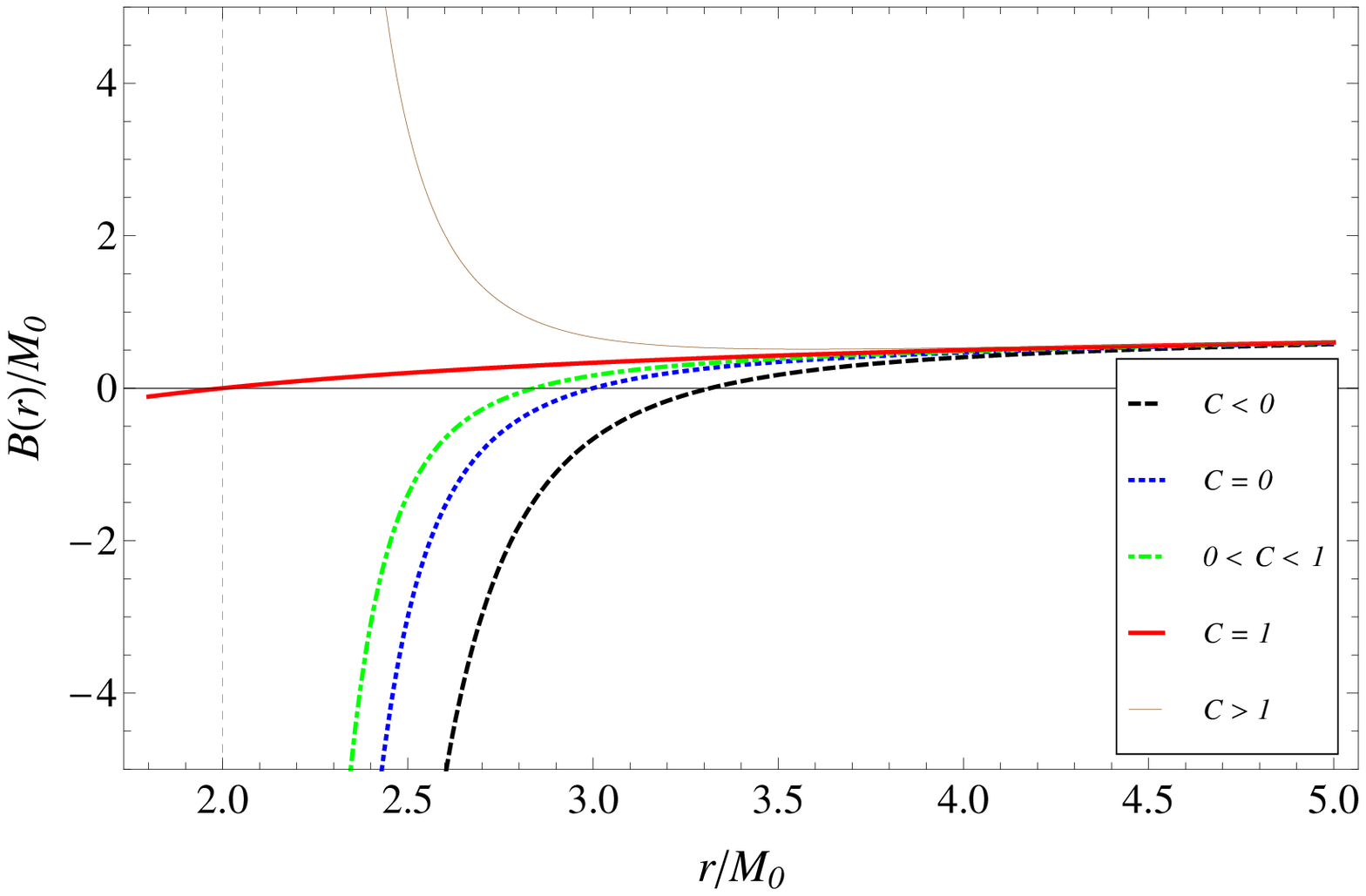}\includegraphics[scale=0.42]{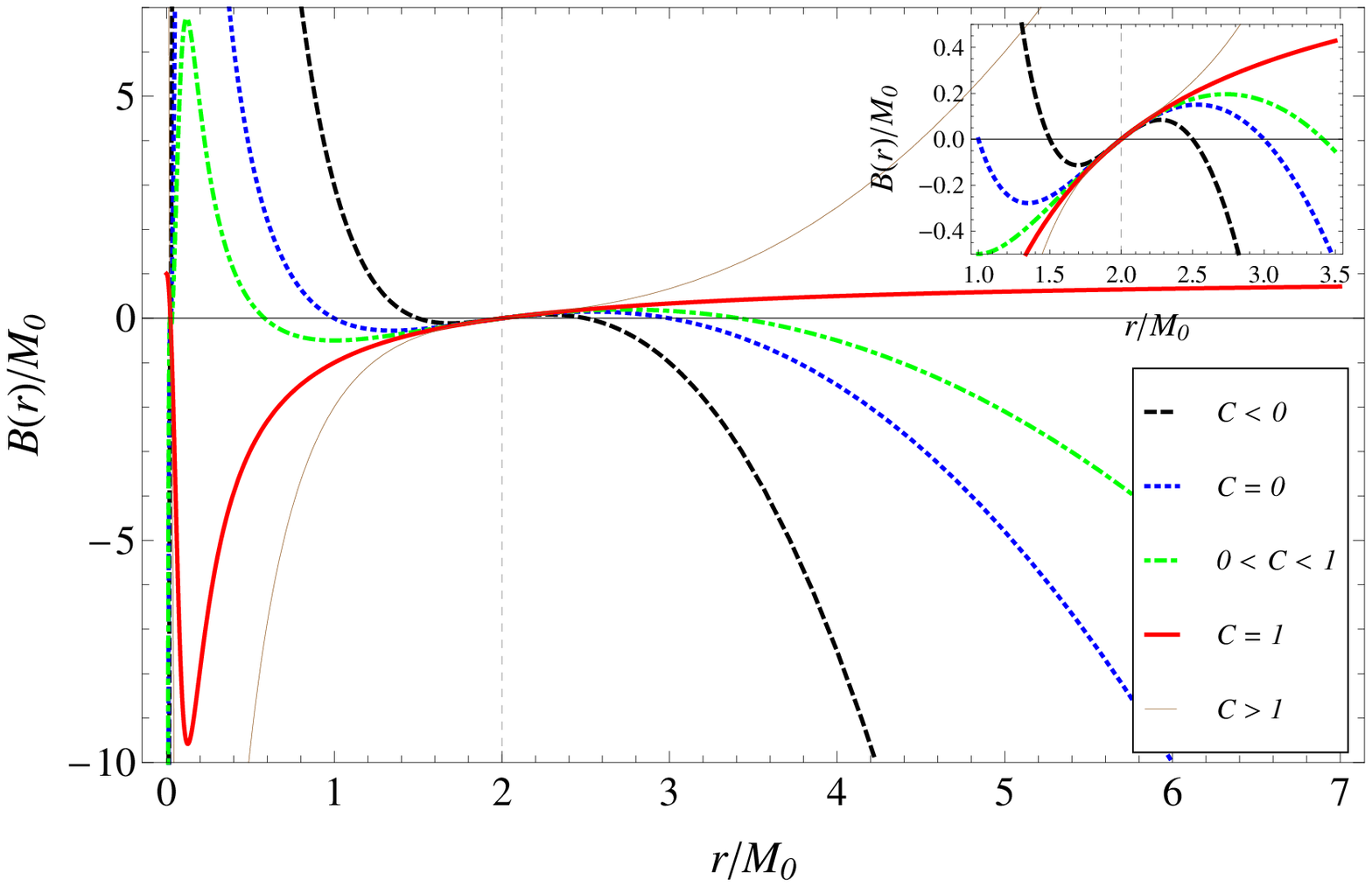}
\par\end{centering}

\caption{Left: $B(r)$ obtained with the mass function (\ref{Mass_term}) and
the constraint $p_{r}=\omega\rho$ for $\omega>0$ or $\omega<-c_{0+}/(1+c_{0+})$.
Right: the same function for $-c_{0+}/(1+c_{0+})\leq\omega<0$. The
vertical dashed line corresponds to the event horizon $r_{+}$. In this graphic
we have used $r_{0}/M_{0}=1,$ $q=3$, $\omega=1/3$ (left), $\omega=-1/3$ (right), $C=-2$, $C=0.5$, and $C=2$.}
\end{figure}

With $-c_{0+}/(1+c_{0+})\leq\omega<0$, the divergence of $B(r)$
and $K(r)$ occurs at $r=r_{0-}$ for the deformed metrics. \textbf{$B_{lin}^{small}(r)$},
by using (\ref{B_small}),\textbf{ }reads
\begin{equation}
B_{lin}^{small}(r)=\frac{A_{0}(r)}{r^{\left(\frac{1+\omega}{\omega}\right)}}\left[\omega(2M_{0}r^{2}-r_{0}^{3})\right]^{\frac{1}{2}\left(\frac{1+\omega}{\omega}\right)}.\label{Blin_worm_small}
\end{equation}
Thus, $r_{0-}=\sqrt{r_{0}^{3}/(2M_{0})}$. In the Fig. 3 (right), by fixing $C>1$, one has a singular BH. In this range of $\omega,$ the values of $C$ which correspond to the physical solutions
with correct metric signature are $C=1$ and $C>1$.

\section{Weak energy condition}

In general, according to Section 2, RBHs with spherical
symmetry do not violate the WEC. The weak violation occurs, as we have
shown in \cite{Neves_Saa}, when the rotation, represented
by metrics with axial symmetry, is present. However, the family of deformed spherical solutions,
obtained from a regular ansatz in this work, violates the WEC. For example, in the $R=0$
case, for $q=3$, the relations among $\rho$ and $p's$ are 
\begin{eqnarray}
\rho+p_{r} & \propto & \left(C-1\right)\frac{\left(r-r_{+}\right)}{r^{2}\left(r-r_{0+}\right)^{2}},\label{rho+pr}\\
\rho+p_{t} & \propto & -\frac{\left(C-1\right)}{r^{2}\left(r-r_{0+}\right)^{2}}+\mathcal{O}\left(1/r^{8}\right).\label{rho+pt}
\end{eqnarray}
For any $C\neq1,$ the WEC is always violated (inside or/and outside of the event horizon $\rho+p_{r}<0$ or/and $\rho+p_{t}<0$).
This conclusion may be reached if $m(r)=m$ is used (the Schwarzschild metric) in the ansatz (\ref{A_0}). 

Therefore, in the specific case $R=0$ with $C_{0}<C<1$, \textit{one has a RBH with spherical symmetry which violates
the WEC}. This interesting result is possible only for deformed metrics, i.e., $A(r)\neq B(r)$.

\section{Cosmological fluids}

It is worth noting that the isotropic case may produce BHs in cosmological
scenarios. Such solutions present one of the most important characteristics
of the modern cosmology: the isotropy at large scales. Assuming (\ref{B}) as $B(r)=A_{0}(r)\left[1+(C-1)b(r)\right]$,
with $b(r)=exp\left[\int k/(rh(r))dr\right]$, one has
\begin{equation}
\omega(r_{+})=\frac{p(r_{+})}{\rho(r_{+})}=-1,\label{omega_horizon}
\end{equation}
with $p=p_{r}=p_{t}$ in this case. According to refs. \cite{Molina2,Martin-Moruno},
which used an ansatz with constant mass, $m(r)=m$, this result indicates
which deformed BHs do not accrete the fluid with equation of state
(\ref{omega_horizon}). Such an equation of state is the same of BHs
in asymptotically de Sitter geometries. And these geometries do not
accrete with such a condition. That is, there is no accretion of the
cosmological constant behaving as a fluid. On the other hand, we have
shown in \cite{Neves_Saa2} that RBHs (non-deformed, i.e., $A(r)=B(r)$) accrete and lost their masses when
the test fluid is phantom type, $\omega<-1$. This result is the same
obtained with singular BHs (see, for example, \cite{Babichev1}).
However, with the mass function (\ref{Mass_term}), the accretion
occurs when 
\begin{equation}
\left(\frac{r_{0}}{r_{c}}\right)^{q}<\frac{1}{2},\label{critical}
\end{equation}
with $r_{c}$ playing the role of a critical point, where the speed of flow is equal
to the speed of sound.

\section{Final remarks}

By employing the approach applied in \cite{Molina2,Casadio,Bronnikov,Various,Various2,Various3} and generalized in \cite{Molina}, we have deformed RBHs for the first time in the literature. This approach works by means of linear constraints in the energy-momentum tensor. From a known ansatz (a determined geometry such as Schwarzschild, Reissner-Nordström, Bardeen or Hayward solution), it is possible to build metrics which represent different types of solutions: regular or singular black holes, wormholes and naked singularities. In this work, our ansatz was given by a regular metric with a general mass function, proposed by us in \cite{Neves_Saa}, to construct RBHs with either spherical or axial symmetry. By using this mass function, the spherical deformed geometries presented here are, in general, singular. That is, the standard RBHs used as ansatz (Bardeen and Hayward solutions) avoid the singularity issue, but the deformed metrics derived from them diverge inside the event horizon. However there is an important exception: with $C_{0}<C<1$ in the brane world context (or, equivalently, a traceless energy-momentum tensor in GR context), one has a deformed RBH. 

Another important issue studied here was the WEC violation. The standard RBHs with spherical symmetry do not
violate the WEC. However, deformed solutions may in general violate this energy condition outside and inside the event horizon, according to eqs. (\ref{rho+pr})-(\ref{rho+pt}). Therefore, \textit{it is possible to obtain a RBH with spherical symmetry (the case $C_{0}<C<1$ above quoted) that violates the WEC}. 

Lastly, in the isotropic case for the metrics presented in this work, the equation of state of the deformed metrics at the event horizon, eq. (\ref{omega_horizon}), indicates the same value when we have an asymptotically de Sitter BH with energy-momentum tensor given by the cosmological constant. These de Sitter-type objects, according to ref. \cite{Martin-Moruno}, do not accrete a fluid represented by the cosmological constant. The same interpretation is possible for deformed metrics presented here. Moreover, as we can see in (\ref{critical}), the accretion gives a constraint for the value of $q$ in the metric.

\acknowledgments
This work was supported by Fundação de Amparo à Pesquisa do Estado
de São Paulo (FAPESP), Brazil (Grant 2013/03798-3). I would like to
thank Alberto Saa for comments and suggestions.


\end{document}